\def\be{\begin{equation}}

\def\bea{\begin{eqnarray}}
\def\nn{\nonumber}
\def\tea{\end{eqnarray}}
\def\hzero{\hbar \rightarrow 0}

\def\d{\delta}

\def\S{\Sigma}

\newskip\humongous \humongous=0pt plus 1000pt minus 1000pt

\newif\ifdtup

\documentstyle[12pt]{article}

\makeatletter                    %allow @ in command names
\@addtoreset{equation}{section}  %reset equation count in each section
\makeatother                     %disallow @ in command names

\begin{document}
\title{Decoherence, Delocalization and Irreversibility
 in Quantum Chaotic Systems}
\author{K. Shiokawa $^{1,3}$ and B. L. Hu $^{1,2,3}$ \\
{\small $^1$ Department of Physics, University of Maryland, College Park,
MD 20742, USA}\\
{\small $^2$ School of Natural Sciences, Institute for Advanced Study,
Princeton, NJ 08540, USA}\\
{\small $^3$ Department of Physics, Hong Kong University of Science and
Technology,} \\
{\small Clear Water Bay, Kowloon, Hong Kong}}
%\date{\today}
\date{\small {\it (IASSNS-HEP-95/21, UMDPP 95-082, March 14, 1995)}}
\maketitle
\begin{abstract}
Decoherence in quantum systems which are classically chaotic is studied.
The Arnold cat map and the quantum kicked rotor are chosen as examples of
linear and nonlinear chaotic systems. The Feynman-Vernon influence functional
formalism is used to study the effect of the environment on the system. It is
well-known that quantum coherence can obliterate many chaotic behavior
in the corresponding classical system. But interaction with an environment can
under general circumstances quickly diminish quantum coherence and reenact
many classical chaotic behavior. How effective decoherence works
to sustain chaos, and how the resultant behavior qualitatively differs from
the quantum picture  depend
on the coupling of the system with the environment and the spectral density and
temperature of the environment.
%Under conditions where the system does not
%decohere completely, such as weak coupling and low temperature, features of
%quantum chaos may persist, such as recurrence and diffusion suppression.
We show how recurrence in the quantum cat map is lost
and classical ergodicity is recovered due to the effect of the environment.
Quantum coherence and diffusion suppression are instrumental to dynamical
localization for the kicked rotor.
 We show how environment-induced effects can destroy this
localization. Such effects can also be understood
as resulting from external noises driving the system.
Peculiar to decohering chaotic systems is the apparent transition from
reversible to irreversible dynamics. We show such transitions
in the quantum cat map and the kicked rotor and distinguish it
from apparent irreversibility originating from dynamical instability
and imprecise measurements. By performing a time reversal on and following the
quantum kicked rotor dynamics numerically, we show how  the otherwise
reversible quantum dynamics acquires an arrow of time upon the
introduction of noise or interaction with an environment.
\end{abstract}

\newpage

\section{Introduction and Summary}

\subsection{Quantum versus Classical Chaos}

The problem of quantum chaos has been intensively studied in the recent decade
\cite{CasaFord,Ford,Ozorio,Gut,Gian,LicLib,Heiss,DX3,Nak,Ott,Ikeda}.
Although the precise criteria for quantum chaos is still
not well established at
this stage, the salient features of a quantized classically-
 chaotic system are better understood than before.
In classical dynamics, chaos appears as the result of instability
caused by nonlinearity  or the compactness of the phase space,
as manifested in the quantum kicked rotor and the Arnold cat map
\cite{LicLib}, two examples we will discuss in this paper.
The degree of instability can be measured by the exponential rate
of separation of initially infinitesimally-close intervals, namely,
the Lyapunov exponent.
 When this local
instability occurs for the entire phase space,
global chaos sets in.
\footnote{Ergodicity thus generated
 is one of the basic criteria for the
validity  of equilibrium statistical mechanics.
Infinite repetition of streching and folding in the phase space may be the
cause for the generation of self- organized structures in the microscopic
world.}

To look for similar phenomena in quantum systems one encounters basic
difficulties. To begin with, the very concept of trajectories
which is used to define classical chaos is meaningless in quantum mechanics.
The equations of motion in quantum mechanics is linear.
Seeking  nonlinear effects in these linear equations, as well as
using concepts of determinacy in a theory based on probabilistic
interpretations
are intrinsically prohibitive.
The words 'quantum chaos' generally refer to possible traces or shadows of
chaos in the quantum  system  obtained  from quantizing the corresponding
classical  system which are known  to possess chaotic behavior.
The study of quantum chaos is devoted to finding
how the classical notion of instability changes when the system is
quantized, and how such  changes can be expressed in the language
of quantum mechanics.
For example, fingerprints of classical chaos may appear as
scars in the wavefunction, as fluctuations in the spectrum, or as
diffusion localization, etc \cite{LicLib},.

How are these classical and quantal characteristics
related to each other in the correspondence
between quantum and classical chaos?
There are many criteria  of classicality,
an issue whose  recent resurgence of interest is stimulated by developments
in many areas  of physics (see, e.g.,  \cite{DX4}). Using the uncertainty
principle as one criterion,  we  see  immediately that there  is
a fundamental discrepancy between the definition of chaos and quantum
uncertainty \cite{HuZha}.
 For systems with conservative dynamics,
the initially infinitesimally-separated trajectories in phase
space will exponentially diverge in some direction and converge
in some other direction. This will soon become incompatible with the quantum
uncertainty principle which prevents one to specify details between  points
in the phase space separated closer than the Planck constant $h$.
Therefore, the fact that many classically-chaotic systems produce
infinitely- folded,  Cantor-set stuctures which can continue to
arbitrarily small scale due to nonlinearity is in conflict with
quantum mechanics.

%Also many chaotic systems studied  so far are deterministic in nature,
%while the quantun mechanics has its basis on the probablistic interpretation.
%Thus, the characteristics of quantum mechanics are refered to the wave and
%statistical properties whose classical limit are highly nontrivial.

In classical mechanics, nonlinearity makes the dynamics sensitive to finer
scales, leading to various fractal structures.
However, in quantum mechanics,
owing to the linear nature of Schr\"{o}dinger's equation, one expects to see
limitations to such fine structures. Quantum effects are known
to smooth out the many-folded trajectories caused by nonlinearity.
In this respect, quantum effect is similar
to the effect of noise on classically chaotic systems \cite{Kap}.
In fact, one can study the scaling property from the classical to the
quantum regimes in a system when $\hzero$ as if the system is subject to
some external noise \cite{Pra}.\\

\subsection{Classicality as an Emergent Behavior of Quantum Open Systems}

The  above description of quantum-classical correspondence takes
the point of view that quantum effect is
a correction to the underlying classical dynamics, which is the attitude
taken by many work on this  subject  using semi-classical approximations.
However, this is opposite to how nature works: most of us would agree  that
quantum mechanics is the fundamental theory which describes nature,
and classical mechanics is only an  approximation to it.

How classical dynamics arises from the fundamental principles of
quantum mechanics
and how our ordinary classical experience can be reconciled with the
quantum depiction
have been the basic questions asked in the foundation of quantum mechanics
and in quantum measurement theory.
Although many different explanations exist, the environment-induced
decoherence point of view seems to be one of simple and practical importance.
\cite{envdec}.
In this point of view,
the quantum to classical transition is induced by
the interaction of a quantum system with an environment.
The averaged effect from coarse-graining the
large number of degrees of freedom is the diminuation of
quantum coherence
and the appearance of diffusion and dissipation in the effective dynamics of
the system.
The decoherence time is defined as
\begin{equation}
 t_{dec} = {1 \over \gamma}(\frac{\lambda_{\theta dB}}{\delta x})^2
\end{equation}
where $ t_{dis}= \gamma^{-1}$ is the dissipation time scale ($\gamma$ is the
damping constant), $\lambda_{\theta dB} = h / \sqrt{2\pi m k T}$
is the thermal de Broglie wavelength,
and  $\delta x$ is the characteristic size of the system
(here we assume a coordinate coupling $x$).
 The decoherence time is usually very short for a bath at high temperatures.
We refer the readers to recent reviews on this topic \cite{envdec}.

This approach has been  applied  to  problems involving quantum  decoherence
in quantum measurement  theory, mesoscopic physic, quantum  cosmology
and semiclassical gravity \cite{DX4}.
 However, The quantum and classical correspondence
of chaotic systems in terms of environment-induced decoherence has so far
been studied only by a limited number of authors \cite{TamSip,ZurPaz,Brun}.

\subsection{Decoherence, Localization and Irreversibility}

\subsubsection{Noise and Localization}

There are many  detailed studies of the classical and quantum kicked rotor
model \cite{LicLib}.
A  particularly interesting feature of quantum nonlinear chaotic systems
is the localization  of wave functions in momentum space
\footnote{  The word "localization" here refers to Anderson localization
\cite{AndLoc},
not to the establishment of delta-functional
correlation between, say, the momentum and the coordinate in the realization
of the quasi-classical state which is sometimes used in the context of
decoherence and quantum to classical transition \cite{Gallis}.}
due to quantum coherence \cite{CCIF}.
This momentum-space localization of the wave function is often compared with
Anderson localization \cite{AndLoc} of electrons in a random potential.
Localization in the kicked rotor is considered
to occur by a similar mechanism \cite{FGP}.
If one views classicality as an emergent behavior of a
decohered quantum system, then it is of interest to  study  the effect  of
an environment on localization.
Dittrich and Graham \cite{DitGra} studied the kicked rotor in a
harmonic oscillator bath, and derived a master equation for the open system.
Their argument is mainly focused on the effect of dissipation
induced by the environment.
They used a low temperature approximation and in the zero temperature
limit they claim that the dynamics becomes Markovian.
This rather unusual behavior is due to the special non-Ohmic
environment they used.
In general, the Markovian regime corresponds only for an ohmic bath at high
temperature.
Cohen and Fishman \cite{CohFis} used the influence functional method
\cite{FeyVer} to study
the effect of noise associated with an Ohmic bath on localization for the
QKR and a similar model.
They calculated explicitly the diffusion constant and the
relevant time scales in terms of the noise correlation and the nonlinear
parameter.
On a related  problem, Ott, Antonsen and Hanson \cite{OAH}
first showed numerically that  external noise breaks the localization of
a wave packet in the QKR. Cohen also studied the effect of noise correlations
\cite{CohFis}. Naively one does not expect correlations to
play an essential role for chaotic systems
because the memory in such systems is lost quickly.
However, in the quantal case, long range correlations  may alter the
situation in a complicated way. In fact it is known that the appearance of
noise autocorrelation depends on the system-environment coupling.

\subsubsection{Irreversibility}

Using  a  simple linear continuous  model, the inverted harmonic
oscillator potential,  Zurek and Paz \cite{ZurPaz}
recently observed that in  the presence of noise, the dynamics can
change from a reversible classical one to an irreversible one.
We  show that a similar behavior exists in the kicked rotor model.
For a conserved Hamiltonian chaotic system, volume conservation causes
one direction in phase space to contract exponentially.
 Without interaction with an environment,
the other source of irreversibility intrinsic to  chaotic
systems arises from the limitation of  actual measurements.
For example, the classical kicked rotor behaves essentially irreversibly
due to the instability of trajectories. However,
in the quantal case, the system characterized by
the quantum state becomes highly stable in spite of the nonlinearity
of the Hamiltonian \cite{Chi}.
When this system interacts with an environment,
there exists a sharp transition from a quantum reversible conservative
stage to a classical irreversible stage.
Irreversibility ascribed to by a limitation of measurement
will be replaced by irreversibility arising from coarse graining
the environment.

\subsection{Time Scales of Competing Processes}

One way to gauge  the relative importance of the pertinent processes
which can influence the dynamics of a  quantum chaotic system is to compare
their characteristic time scales. Let us start with cases  with no
interaction with an environment.
There are essentially two different time scales involved.
% to characterize  the evolution of the system without the bath.
One is the Ehrenfest time $t_E$ and the other is the relaxation time $t_R$.
The Ehrenfest time $t_E$ is defined as the time
within which the Ehrenfest theorem holds.
\begin{equation}
t_E \sim \frac{1}{\lambda} ln \frac{\delta p(0)}{\hbar},
\end{equation}
where $\delta p(0)$ is the relevant initial (angular) momentum scale.
Violation of the Ehrenfest theorem
in the quantal case arises from the nonlinear terms in the potential
which can be seen in the evolution (Kramer-Moyal)
equation of the Wigner function.
\begin{eqnarray}
 \frac{\partial W(X,p)}{\partial t}
&=&
- \frac{2}{\hbar} H \sin
(\frac{\hbar}{2}
( \frac{ \stackrel{\leftarrow}{\partial} }{\partial p}
 \frac{ \stackrel{\rightarrow}{\partial} }{\partial X}
 - \frac{ \stackrel{\leftarrow}{\partial} }{\partial X}
 \frac{ \stackrel{\rightarrow}{\partial} }{\partial p}) ) W(X,p)     \\
&=&
\{ H, W \} + \Sigma_{n=1}^{\infty} (-1)^{n}
 (\frac{2}{\hbar})^{2n}  \frac{1}{(2n+1)!}
\frac{\partial^{2n+1} H}{{\partial X}^{2n+1}}
\frac{\partial^{2n+1} W}{{\partial p}^{2n+1}}
\end{eqnarray}
where $\{ ~ \}$ is the Poisson bracket.

The appearance of localization arises from the discrete spectrum of the
Hamiltonian.
Thus the time it takes for the wave packet to localize is determined
by how long it takes for the system to recognize the discreteness of the
spectrum.
Simple argument is given in \cite{CCIF}:
Since $l$ represents the effective number of modes scattered in the period
$[0,2\pi]$, the typical spacing is given by $\Delta \omega \sim 2\pi /l$.
Thus after $t_R \sim 1/\Delta \omega \sim l$,
the system localizes. (See also discussion in Sec. 3.3.)

Upon interaction with a bath, a system effectively decoheres  at
the decoherence time scale $t_{dec}$.
Another time scale  $t_C$ arising from the coarse graining
appears which also contributes to the
violation of the Ehrenfest theorem. As discussed in \cite{ZurPaz},
it determines the transition regime from the reversible
 classical Liouville dynamics to the irreversible dynamics as embodied in
 the Second Law of Thermodynamics.

 From our study, this picture also holds for the kicked rotor.
 In this case, we see the transition from the initial
 constant-entropy regime to the entropy-increasing regime.
 Because of the compactness of the space, we see entropy does not
 increase forever but will eventually saturate.
 After $t_C$, the evolution is not unitary.
Note that even if the evolution of the Wigner function is the same
 as that of the classical Liouville distribution function before $t_E$,
 one  should not regard the system as in a classical state.
\footnote{Note that,  as shown in the examples of  \cite{BYZ94},
the Ehrenfest theorem is neither necessary nor  sufficient to define
classicality. There are systems which evolve strictly quantum mechanically
but the expectation
values of the  canonical variables obey classical equations;
and there are models which do not satisfy the theorem
but their evolution is essentially classical.}

Dynamical localization is completed at the relaxation time $t_R$.
 At decoherence time $t_{dec}$, coherence is destroyed up to the localization
 length. If $t_{dec} >> t_R$, suppression of momentum diffusion due to
quantum effects always exists and we will never see the classical state.

 It is known that the evolution of the kicked rotor is not time reversible
 while its quantized version is time reversible.
 This type of quantum stability is also considered to be one of
 the characteristics of quantum chaos. In a quantum system,
 irreversibility arising from
 limited precision in a measurement now no longer causes serious loss
 of information.   Instead, interaction with a bath introduces the
 irreversibility due to the coarse graining for the quantal case.

In this paper, we study the quantum dynamics of two simple
 models which possess classical chaotic behavior,
the Arnold cat map and the kicked rotor.
By introducing linear coupling with a harmonic oscillator bath
assumed to be Ohmic and at high temperature,
we show how the effective dynamics of a quantum open system reveals
the well-known classical chaotic behavior.
In Section 2, we examine the quantum cat map (QCM) of a system
coupled with a harmonic bath.
The system is known to be chaotic when the corresponding matrix for the
map is hyperbolic.
We use the influential functional method to
study the effect of the environment on this system.
	By measuring the linearized entropy
we show that the decoherence mechanism works more efficiently than
the regular case. Namely, the rate of decoherence is faster in
the chaotic system. Decoherence rate in chaotic systems was also studied by
Tameshit and Sipe \cite{TamSip}.
Peculiar to the quantum case is the recurrence behavior of physical
quantities, resulting from the finiteness of the phase space points
in the quantum map due to the quantization (
because the phase space is periodic in both the coordinate and momentum).
We show that interaction with the environment erases the recurrence
in the hyperbolic map but not in the elliptic map.
Thus both cases behave close to the correponding classical limit.

In Section 3, we examine the quantum kicked rotor (QKR)
as a prototype of
nonlinear chaotic systems.
Without interaction with a bath,
the wave function shows localization arising from quantum coherence
effects.
 Loss of coherence due to
 interactions with an environment shown by the decay of
 the off-diagonal components of a reduced density matrix is responsible
for the breaking of
 localization.
 The decay rate increases as the noise strength associated with the
environment and the nonlinear parameter get larger.
 In Section 4, we examine the transition from reversible to irreversible
 dynamics intrinsic to an unstable system
 due to the interaction with an environment.
 We show that the same mechanism holds for both the cat map and kicked
 rotor.
 For both cases, the entropy shows saturation
 possibly due to the bounded nature of the phase space.
 Furthermore, we perform time-reversal numerically and
show how the interaction with an environment changes
the nature of irreversability.
Details of results can be found at the end of each section.

\section{Decoherence in a Linear Map}
\setcounter{equation}{0}

\subsection{Quantum Cat Map}

The cat map is a linear area-preserving map $T$
on a two-torus in phase space
by identifying the boundaries of the interval $[0,2 \pi]$
in both the coordinate $Q$ and the momentum $P$ directions \cite{ArnAve}.
 From time step $j$ to $j+1$, it is given by
\begin{equation}
		\left( \begin{array}{c}
		Q_{j+1}\\
		P_{j+1}
		 \end{array}      \right)  =
		\left( \begin{array}{cc}
		a & b \\
		c & d
		 \end{array}      \right)
		\left( \begin{array}{c}
		Q_{j}\\
		P_{j}
		 \end{array}      \right)  =  T
		 \left( \begin{array}{c}
		Q_{j}\\
		P_{j}
		 \end{array}      \right)
\end{equation}
where $det T = 1$ guarantees area preservation.
The degree of chaos depends on the choice of $T$.
The eigenvalues of $T$ are either both real or both imaginary.
In the latter case, $T$ is elliptic, the motion becomes periodic
and no sensitive dependence on the initial
condition is observed.
When $T$ is hyperbolic,
the motion becomes chaotic.

Quantized cat map is studied in detail by Hannay and Berry \cite{HanBer}.
Due to the
periodicity and thus the discreteness of both phase space variables
, the area of the torus is characterized by a discrete
 Planck's constant,
\begin{equation}
		 \hbar = 2 \pi / {\cal N}
\end{equation}
where ${\cal N}$ is the number of sites in both the coordinate and the momentum
directions in phase space.
Because of this, quantum dynamics
defined by the cat map is considered to describe quantum resonance.
Note that this is not a generic feature for other systems
which have continuous phase space.

The action $S(Q_{j+1},Q_{j})$ for this linear map is easily constructed
from conditions
\begin{equation}
	\frac{\partial  S(Q_{j+1},Q_{j})}{\partial Q_{j+1}} = P_{j+1}; ~~~
	 -\frac{\partial  S(Q_{j+1},Q_{j})}{\partial Q_{j}} = P_{j}
\end{equation}
Combining (2.1) and (2.3) gives
\begin{equation}
	 S(Q_{j+1},Q_{j}) = \frac{1}{2 b} ( a Q_{j}^2 - 2 Q_{j} Q_{j+1}
							 + d Q_{j+1}^2 )
\end{equation}
Before imposing the periodic boundary conditions, the propagator is
\begin{equation}
 U(Q_{j+1},Q_{j}) = \frac{1}{2\pi}(\frac{i{\cal N}}{b})^{\frac{1}{2}}
\exp{\frac{i{\cal N}}{4\pi b}( a Q_{j}^2 - 2 Q_{j} Q_{j+1} + d Q_{j+1}^2 )}
\end{equation}
With periodic boundary conditions,
one needs to sum over all equivalent initial
points, thus yielding
\begin{eqnarray}
	 U(Q_{j+1},Q_{j})
		& = & \frac{1}{2\pi}(\frac{i{\cal N}}{b})^{\frac{1}{2}}
		\sum_{m=-\infty}^{\infty}
	 \exp[   \frac{i{\cal N}}{4\pi b}
( a (Q_{j}+2\pi m)^2 - 2 (Q_{j}+2\pi m) Q_{j+1} + d Q_{j+1}^2 ) ] \nn \\
		& = &
	 C(T,{\cal N}) \exp[ \frac{i{\cal N}}{4 \pi b}
		( a Q_{j}^2 - 2 Q_{j} Q_{j+1} + d Q_{j+1}^2 ) ]
\end{eqnarray}
where $C(T,\cal N)$ is a constant depending on the form of $T$ and $\cal N$.

In fact, $C(T,\cal N)$ vanishes in many choices of $T$ and
this sum gives a nontrivial value for the progagator only if the matrix has
a special form.
We choose
\begin{equation}
		 T_{1} = \left( \begin{array}{cc}
		 0 &  1 \\
		-1 &  0
		 \end{array}      \right),
\end{equation}
for the elliptic case and
\begin{equation}
		 T_{2} = \left( \begin{array}{cc}
		 2 & 1\\
		 3 & 2
		 \end{array}      \right),
\end{equation}
for the hyperbolic case.

For the special choice of the matrix elements $T_{1}$ and $T_{2}$ made above,
the propagator takes on the simple form,
\begin{equation}
U_1(j+1, j) = \sqrt{\frac{i}{\cal N}}\exp[-\frac{i}{\hbar}Q_{j}Q_{j+1}],
\end{equation}
\begin{equation}
U_2(j+1, j) = \sqrt{\frac{i}{\cal N}}\exp[ \frac{i}{\hbar}
(Q_{j}^{2}- Q_{j}Q_{j+1}+Q_{j+1}^{2})].
\end{equation}
Since each iteration describes a permutation among sites,
each site belongs to a periodic orbit.
Thus quantum evolution follows the classical dynamics,
resulting in the recurrence of the wave function (or equivalently,
of the Wigner function) \cite{HanBer}.

\subsection {Decoherence in the Quantum Cat Map}

We now couple the system to a bath of $N$ harmonic oscillators linearly.
\footnote{ Nonlinearity in the coupling could enhance the interaction
 between the system and the bath.
 One mode in the system couples with
 numbers of different modes in the bath in the presence of nonlinear
 coupling.
 Then this type of coupling effectively increases the number of system in the
 bath $N$.
 Consequently, one can expect it helps decohere the system.
 Nonlinearity in the bath seems to have a similar effect.
 \cite{Sak}}.
 The Hamiltonian of the bath of oscillators with coordinates $q_\alpha$
 and momentum $p_\alpha (\alpha = 1,..,N)$ is
\begin{equation}
 H_{B} = \sum_{\alpha=1}^{N}( \frac{p_{\alpha}^{2}}{2}
			+ \frac{\omega_{\alpha}^{2}q_{\alpha}^{2}}{2}).
\end{equation}
The interaction Hamiltonian between the system $Q$ and the bath
 variables $q_\alpha$
is assumed to be bilinear,
\begin{equation}
 H_{C} = \sum_{\alpha=1}^{N} C_{\alpha} Q q_{\alpha}.
\end{equation}
where $C_\alpha$ is the coupling constant of the $\alpha$th oscillator.

By integrating out the bath variables, we get the reduced density matrix,
\begin{equation}
	\rho_r(Q_{j}, Q'_{j}, t) =
	\int \Pi_{\alpha=1}^{N} dq_{\alpha} dq'_{\alpha} \exp \frac{i}{\hbar}
			[S(Q)+S_C(Q,q_{\alpha})+S_B(q_{\alpha})
			-S(Q')-S_C(Q',q'_{\alpha})- S_B(q'_{\alpha})].
\end{equation}
where $S$ is the classical action of the system defined in (2.4)
and $S_B$, and $S_C$ are the actions for the bath and the coupling
respectively.
The evolutionary operator $J_r$ for the reduced density matrix
from time steps $j$ to $j+1$ is
\begin{equation}
	J_r(Q_{j+1}, Q'_{j+1} \mid Q_{j}, Q'_{j}, t) =
			\int DQ DQ' \exp \frac{i}{\hbar}[S(Q)-S(Q')+ A(Q, Q')]
\end{equation}
in a path-integral representation \cite{FeyVer,CalLeg83,Gra,HPZ}, where
\begin{equation}
\frac{i}{\hbar}A(Q, Q')= \frac{1}{\hbar^{2}}\int_{0}^{t} ds \int_{0}^{s} ds'
 r(s)[-i \mu(s-s')R(s') - \nu(s-s') r(s')]
\end{equation}
is the influence action.
Here  $ r \equiv \frac{1}{2}(Q - Q'),
 R \equiv \frac{1}{2}(Q + Q')$, and $\mu(s), \nu(s)$ are
the dissipation and noise kernels respectively \cite{HPZ}.

If we consider the simplest case  of an ohmic bath at high temperature
 $kT > \hbar \Lambda >> \hbar \omega_\alpha$ \cite{CalLeg83},
 and consider times shorter than the dissipation time,
then we obtain a Gaussian form for the influence functional, with
$    {\frac{i}{\hbar}A(Q, Q')} =
		 {-\frac{2 M \gamma k T}{\hbar^{2}} \Sigma_{j} r_{j}^{2}}.
$
where the noise kernel becomes local $\nu(s) = 2 M \gamma k T \delta(s)$
and $\gamma$ is the damping coefficient.
\footnote{ Because the chaotic trajectory washes out information
 about the past rapidly, we expect that memory effect would be less important
in classically chaotic systems than classically regular systems.
Nevertheless, in the quantal case in which the classical
stretching and folding behavior is suppressed, it would be still interesting
to study how the non-Markovian behavior competes with nonlinearity
\cite{CohFis}.
We will discuss this issue in the next section.}
The unit-time propagator becomes
\begin{equation}
	J_{r}(Q_{j+1}, Q'_{j+1} \mid Q_{j}, Q'_{j}) = \langle
	J_{r}(Q_{j+1}, Q'_{j+1} \mid Q_{j}, Q'_{j}, \xi)  \rangle
 =\langle \exp \frac{i}{\hbar} [S(Q_{j+1},Q_{j})- S(Q'_{j+1},Q'_{j})
	+2 \xi r_{j+1}] \rangle.
\end{equation}
Here $\xi$ is a Gaussian white noise given by
\begin{equation}
 \langle \xi \rangle = 0 , ~~~
 \langle \exp \frac{2i}{\hbar} \xi r \rangle
 = \exp [   -\frac {2 M \gamma k T}{\hbar^{2}} r^2 ]
\end{equation}
where $\langle ~\rangle$ denotes statistical average over noise realization
$\xi$.

For the elliptic map, we get
\begin{equation}
	J_{r}(Q_{j+1}, Q'_{j+1} \mid Q_{j}, Q'_{j}, \xi)
 = (\frac{i}{\cal N})^{1/2}
 \exp [\frac{2i}{\hbar}  (- r_{j} R_{j+1} - r_{j+1} R_{j} + \xi r_{j+1})].
\end{equation}
and for the hyperbolic map,
\begin{equation}
	J_{r}(Q_{j+1}, Q'_{j+1} \mid Q_{j}, Q'_{j}, \xi)
 = (\frac{i}{\cal N})^{1/2}
 \exp [ \frac{2i}{\hbar}
				(2 r_{j} R_{j} + 2 r_{j+1} R_{j+1}
				- r_{j} R_{j+1} - r_{j+1} R_{j} + \xi r_{j+1} )]
\end{equation}

The Wigner function is defined as
\begin{equation}
 \begin{array}{l}
	W(R,p)
	= \frac{1}{\pi \hbar}
		\int_{-\infty}^{\infty} \psi (R+r) \psi^{*} (R-r)
		\exp[   \frac{-2 i p r}{\hbar}     ] dr.
		\end{array}
\end{equation}
where $p$ is the momentum conjugate to $r$.
The propagator $K_{T}$ for the Wigner function is
\begin{equation}
	\begin{array}{l}
		K_{T}(R_{j+1}, p_{j+1}\mid R_{j}, p_{j}, \xi)
	 =  \frac{1}{\pi \hbar} \Sigma_{r_{j}} \Sigma_{r_{j+1}}
		 J_{r}(Q_{j+1}, Q'_{j+1} \mid Q_{j}, Q'_{j}, \xi)
	\exp  \frac{2 i}{\hbar} ( p_{j} r_{j} - p_{j+1} r_{j+1}).
	\end{array}
\end{equation}
This is reduced to the form of the classical cat map.
For the elliptic case,
 \begin{equation}
R_{j} = -p_{j+1} + \xi, ~~~ p_{j} =  R_{j+1} .
\end{equation}
For the hyperbolic case,
\begin{equation}
R_{j} = 2R_{j+1} -p_{j+1} + \xi , ~~~ p_{j} = - 3R_{j+1} + 2p_{j+1} -2\xi.
\end{equation}

Thus, without the environment,
quantum evolution follows classical permutation \cite{HanBer}.
We can also say that the transformation
 from the classical
map to the corresponding quantum propagator
 $T \rightarrow K_{T}(R_{j+1}, p_{j+1} \mid R_{j}, p_{j}, \xi=0)$
 preserves the group structure.
When coupled to the bath, the cat map
is exposed to a Gaussian
noise from the environment in each time step.
The phase space is divided by a finite number of different
periodic orbits and the
 period is known to increase roughly proportional to ${\cal N}$.
The discretized noise induces transition between different
periodic orbits in an irregular way.
As a consequence, the recurrence of some physical quantity
will disappear in the quantum map and the classical
type of mixing is regained.

Fig.1 shows $Tr \rho_r^{2}$, the linearized entropy (with the
reversed sign) for various cases.
If there is no interaction with the environment,
the entropy is constant for both the regular and chaotic cases.
When interaction sets in, $Tr \rho_r^{2}$ decays
exponentially, showing that the system rapidly decoheres.
There is no recurrence of this quantity observed.
In spite of the discreteness of the points on the torus,
 we expect that the system behaves classically due to the influence of
the environment.

Note that the system also decoheres
in a similar manner when the system is regular but with a slower
 rate \cite{TamSip}.
These results indicate that if the underlying classical system shows
chaotic behavior, even after quantized, the system still posseses the mixing
behavior.
This mixing property enhances
the random perturbations from the environment, thus accelerating the
suppression of quantum interference.
However, in this particular example, the dynamics is
essentially classical as is seen in (2.24)
(In this case, the value of $\hbar$ comes in through the number
 of sites ${\cal N}$).
 More general cases should be examined.

In Fig.2, we show the mean displacement of points in the phase space
from the initial configuration as a function of time.
 This is defined by
$l = \sqrt{\langle \Delta x^2 + \Delta p^2 \rangle}$, where $\Delta x$ and
$\Delta p$ are the displacement from the initial phase space points,
 $\langle~\rangle$ is the average
over the phase space points and noise realizations.
In the chaotic case, we see the recurrence disappears with just a small
amount of noise (Fig.2a).
Whereas in the regular case, the same amount of noise does not alter
the qualitative picture of recurrence (Fig.2b).
In both cases, the decohered quantum system behaves close to the classical
picture in which
the regular and chaotic dynamics are clearly distinguished.
For the elliptic map, the classical dynamics is completely periodic.
For the choice of $T_1$ in (2.7), the period is four.
On the other hand, for the hyperbolic map $T_2$ in (2.8), the classical
dynamics is nonperiodic.
In spite of the discreteness of the points on the torus,
the system  behaves effectively classically due to the effect of
the environment.

%In Fig.3, we show the evolution of the Wigner function.
%Initially, Gaussian-shaped wave packets start deforming.
%For an isolated system (Fig.3a),
%these wave packets will eventually split into tiny pieces
%due to the discrete nature
%of the phase points.
%In the classical case, each packet would stretch infinitely in one direction
%and contract in the other direction.
%This is the source of chaos and fractal structure.
% In the quantum case, quantization admits only a finite
%number of points in the phase space,
% consequently, quantization rule itself prevents the system
%to behave chaotically. However, interaction with a bath delays
%the wave packet splitting  until a much later time (Fig.3b)
%(For a related argument, see \cite{TAI}).

\section{Nonlinearity and Decoherence}
\setcounter{equation}{0}

\subsection{Quantum Kicked Rotor}

The kicked rotor and its map version, known as the standard map,
are one of the most
intensively studied models from both the quantum and classical points of view
\cite{LicLib}.
 The Hamiltonian of the kicked rotor is given by
\begin{equation}
H = \frac{p^{2}}{2m}  + K cos x \Sigma_{j=-\infty}^{\infty} \delta(t - j)
\end{equation}
which describes  a one-dimensional rotor subjected to a delta-functional
periodic kick at $t=j$.
Here $x$ is the angle of the rotor with period $2 \pi$,
$m$ is the momentum of inertia, $p$ is an angular momentum,
 $K$ is the strength of the kick
and, in this case, the nonlinear parameter.
When $K > K_c = 0.9716$, the system becomes chaotic over
the entire phase space.
The average energy $p^2$ is known to show diffusive behavior
like that of a Brownian particle under a stochastic force.
This suggests the emergence of randomness in a
deterministic chaotic system.

The quantum dynamics of the kicked rotor is depicted by the corresponding
Schr\"{o}dinger equation
\begin{equation}
		 i \hbar \frac{\partial}{\partial t}
 \psi (x,t) = -\frac{\hbar^{2}}{2 m}\frac{\partial^{2}}{\partial x^{2}}
 \psi (x,t) + K \cos x \S \d (t-j) \psi (x,t)
\end{equation}
where $\psi$ is the wave function for the rotor.
Denoting $\psi_{j}$ as the wave function $\psi(x,t)$ at each discrete
 time $t = j$,
and integrating (3.2) from $j$ to $j+1$, we obtain
\begin{equation}
	\psi_{j+1} (x) =
 \exp [-i \frac{\hbar}{2m}  \frac{\partial^2}{\partial x^2}]
 \exp [-i \frac{K\cos x}{\hbar} ]
 \psi_{j}(x)
\end{equation}

The quantum kicked rotor~(QKR) is known to exhibit
dynamical localization.
After some relaxation time scale,
the wave function becomes exponentially localized in the momentum space
 \cite{CCIF}.
This may be interpreted as a particle moving in a lattice with
a quasi-random
potential.
This heutistic picture seems to justify the analogy between
the quantum kicked rotor to the tight binding model
with an exponentially decaying hopping parameter which is known to show
 Anderson localization.
The explicit transformation to the tight binding model
was constructed in \cite{FGP}.
In spite of the nonrandom, deterministic nature of the kicked rotor
Hamiltonian, numerical results there assuming sufficient
quasi-randomness
 seem to support this analogy.
Dynamical localization in this context arises from
 the suppression
of classical diffusive behavior
in the quantum dynamics.
As shown by Ott et.al. \cite{OAH}, a small external noise
can break the localization.
Numerically they observed
three different regimes in the behavior of the diffusion constant $D$
depending on the magnitude of noise.
When the noise becomes sufficiently large,
the system recovers the classical diffusive
behavior.

It is of interest to interpret the above results due to noise
 from the microscopic open system
point of view.
When the system interacts with an environment,
we know that coarse graining of the environmental variables
is also a source of noise and dissipation.
We shall now derive the influence functional for the quantum kicked
rotor and study its behavior.

\subsection {Quantum Kicked Rotor in a Bath}

Cohen and Fishman studied the case for the ohmic bath
in detail \cite{CohFis}.
They calculated explicitly the diffusion constant and the
relevant time scales in terms of the noise correlation and the nonlinear
parameter.

 We introduce a linear coupling of the system momentum $p$
 with each oscillator coordinate
 $q_ \alpha (\alpha = 1,..N)$ in the bath in the form
 $ H_C = \Sigma_{\alpha=1}^N C_\alpha q_{\alpha} p $
(Here $q, p$ without the subscript $\alpha$ denote the system coordinate and
 momentum variables).
 For an ohmic environment, the action functional has
 the same form as (2.15), except that the coordinate variable $Q$
 is replaced by
 a momentum variable $p$.
\begin{equation}
\frac{i}{\hbar}A(p, p')  = \frac{1}{\hbar^{2}} \int_{0}^{t} ds \int_{0}^{s} ds'
	p^{-}(s)[-i \mu(s-s')p^{+}(s') - \nu(s-s')p^{-}(s')]
\end{equation}
where $p^{\pm}(s) = p(s) \pm p'(s)$.

In a similar way, we introduce the noise $\xi(\tau)$ such that,
\begin{equation}
	\langle \exp[-i\int \xi(\tau) p(\tau)] \rangle
	= \exp \left[ -{1\over\hbar}\int\limits_0^tds\int\limits_0^{s}ds'
				 p^{-}(s)\nu(s-s') p^{-}(s') \right]
\end{equation}
As before, we will examine processes in the time span where dissipation is
small, thus ignoring the effect of the dissipation kernel $\mu(s)$.

Under these assumptions,
the action of the noise kernel can be formally absorbed in the propagator
for the wave function.
The unit time propagator for the wave function
$ U_\xi(j+1,~j) $ is given by
\begin{equation}
 U_\xi(j+1,~j)
=   \exp[-\frac{iK\cos x}{\hbar}]
		\exp[-\frac{i p^2}{2m} ]
		\exp[-\frac{i \xi p}{\hbar}]
\end{equation}
where, as before, the noise term $\xi $
arises from using a Gaussian
identity in the integral transform of the term involving the noise kernel
in the influence functional \cite{FeyVer}.
Summing over all noise realizations $\langle ~\rangle $
gives the desired reduced density matrix,
\begin{equation}
\rho_{j}(p,~p')
=\langle \psi_{j,\xi}(p) ~\psi^{*}_{j,\xi}(p') \rangle
\end{equation}
where
\begin{equation}
{\psi_{j+1,}}_\xi(p) = U_\xi(j+1,~j) ~{\psi_{j,}}_\xi(p)
\end{equation}
and ${\psi_{j,}}_\xi(p)$ is the wave function under the influence of
a particular noise history represented by $\xi$.

 In the same way, $Tr \rho_r^{2}$ can be expressed as
\begin{equation}
Tr \rho_r^{2} =
 \langle
	 \Sigma_{p} \Sigma_{p'} \psi_\xi(p) ~\psi^{*}_\xi(p')
		~\psi_{\xi'}(p') ~\psi^{*}_{\xi'}(p)
	\rangle_{\xi,\xi'}
\end{equation}
where $ \langle ~\rangle_{\xi,\xi'} $
 denote the statistical average of all possible noise histories
of two independent noises $ \xi(\tau),\xi'(\tau)$ defined at each interval
from $j$ to $j+1$.
At high temperatures,
$\xi(\tau),\xi'(\tau)$ are reduced to two time-uncorrelated
independent Gaussian white noises
defined at each time step.

There are many possible ways of introducing an interaction
 between the system and the bath,though many of them are
related to each other
as shown in \cite{CohFis}.
One interesting case is when we intoroduce the coupling
through the coodinate $x$.
Then to preserve periodicity of the Hamiltonian under the coordinate
transformation $x \rightarrow x + 2 \pi$, we need to restrict the range
of noise to $\xi = n \hbar (n = 0,\pm 1, \pm 2, ..)$, or choose the
interaction Hamiltonian to be $ H_C = C_\alpha q_{\alpha} \cos(x)$.
In the latter case,  we may further assume the
form $ H_C = C_\alpha q_{\alpha} \cos(x + \phi_{\alpha})$,
 where $\phi_{\alpha}$ is the
random phase \cite{OAH} to remove the coordinate dependence
 of the interaction.
 However, they all give the same result, but with different noise
 correlations.
 For example, from (3.6) we can calculate the propagator
 $ U_\xi(j,~1) $ from $t = 1 $ to $t = j$ as
\begin{eqnarray}
 U_\xi(j,~1)
 =&   U_\xi(j,~j-1)~U_\xi(j-1,~j-2)... U_\xi(2,~1)  \hspace{3cm}  \nn    \\
 =&   \exp[  -\frac{iK\cos x}{\hbar}]
	\exp[  -\frac{ip^2}{2m} ]
	\exp[  -\frac{i\xi(j)p}{\hbar}]
	 \times    ....      \hspace{6cm}                            \\
	&    \times
	\exp[  -\frac{iK\cos x}{\hbar}]
	\exp[  -\frac{i p^2}{2m} ]
	\exp[  -\frac{i \xi(1) p}{\hbar} ]               \nn        \\
 =  &
	\exp[  -\frac{i \eta(j) p}{\hbar} ]
	\exp[  -\frac{iK\cos (x + \eta(j))}{\hbar}]
	\exp[  -\frac{i p^2}{2m} ]
	\times   ...   \times
	\exp[  -\frac{iK\cos (x + \eta(1))}{\hbar}]
	\exp[   -\frac{i p^2}{2m} ]                             \nn
\end{eqnarray}
 where   $\eta(j) = \xi(j) +...+ \xi(1)$.
 Thus, this describes the same dynamics as couplings through
 $x$ via $ H_C = C_\alpha q_{\alpha} \sin x$,
 as long as the noise $\eta(j)$ remains small.
 The correlation of the two different noises
 $\eta(j)$ and $\xi(j)$ are related
 by
 \begin{equation}
\langle
\eta(\tau) \eta(\tau')
\rangle
=
\Sigma_{t=1}^{\tau} \Sigma_{t'=1}^{\tau '}
\langle
\xi(t) \xi(t')
\rangle
\end{equation}
Note that even if $\xi$ is a white noise, $\eta$ is not necessarily white.

\subsection {Localization and Decoherence in the Quantum Kicked Rotor}

The eigenvalue equation for
the quasi-energy state in the QKR is given by \cite{FGP}
\begin{equation}
 \exp(-\frac{i}{\hbar}K\cos x) \exp(-\frac{i}{\hbar}\frac{p^2}{2m})
						 u_{\omega}(x)
	= \exp(-\frac{i}{\hbar}\omega)  u_{\omega}(x).
\end{equation}
This can be transformed to
\begin{equation}
\left \{  \exp[ -\frac{i}{\hbar}(\frac{p^2}{2m}-\omega) ]
[ 1 - i \tan( \frac{K\cos x}{2 \hbar} ) ]   -
[ 1 + i \tan( \frac{K\cos x}{2 \hbar} ) ]  \right \}
[ 1 + \exp(-\frac{i}{\hbar}K\cos x)] \frac{1}{2} = 0.
\end{equation}
If we define $   \bar{u}_{\omega}(x)  $ as
\begin{equation}
	 \bar{u}_{\omega}(x)
	 =    [ 1 + \exp(-\frac{i}{\hbar}K\cos x) ] \frac{1}{2}
				 u_{\omega}(x)
\end{equation}
then (3.15) can be written as
\begin{equation}
	\left\{ i\frac{ [ 1 - \exp( -\frac{i}{\hbar}(\frac{p^2}{2m}-\omega) )]}
	{[ 1 + \exp( -\frac{i}{\hbar}(\frac{p^2}{2m}-\omega) )    ]
	- \tan( \frac{K\cos x}{2 \hbar} ) }    \right\} \bar{u}_{\omega}(x) = 0.
\end{equation}
Expanding $\bar{u}_{\omega}(x)$ as $\bar{u}_{\omega}(x)
= \Sigma_{k=-\infty}^{k=\infty}   \bar{u}_k e^{ikx} $, we get
\begin{equation}
	T_k \bar{u}_k + \Sigma_{r\neq0} W_r \bar{u}_{k+r} = E \bar{u}_k
\end{equation}
where
\begin{equation}
	T_k = tan( \frac{\omega}{2 \hbar} -  \frac{\hbar^{n-1} k^n}{4} )
\end{equation}
\begin{equation}
	W_k = \frac{1}{2 \pi} \int_{0}^{2 \pi} d \theta e^{i k \theta}
	tan( \frac{K\cos x}{2 \hbar} )
\end{equation}
and $E = -W_0$.
As pointed out in \cite{FGP}, (3.16) gives the eigenvalue equation
for the tight binding model which
describes electron motion in a quasi-random potential
$T_k$ with the hopping parameter $W_k$.
The property of this model depends on the
 rationality of the coefficient of $k$ in the potential.
When the coefficient is irrational,
the model is known to show Anderson localization.

We can now analyze the relation between the breaking of
 dynamical localization and quantum decoherence.
Loss of quantum coherence is measured by the density matrix
becoming approximately diagonal.
Decoherence in the quantum Brownian model has been
studied extensively for this problem.
We refer the reader to recent work on this topic \cite{envdec}.
In Fig.3a we plot the linearized entropy $Tr \rho_r^{2}$ versus
 the energy $\langle p^2/2 \rangle$ in each diagram
 (we set the mass $m = 1$ for all the numerical calculations).
Note that the two effects are correlated to each other as expected.
This shows that delocalization occurs as the quantum coherence breaks down,
suggesting that delocalization and decoherence occur by the same mechanism.
As the noise strength increases, we see that
 decoherence works more efficiently.
Also as the nonlinearity parameter $K$, the strength of the kick in this model,
increases, the system decoheres more rapidly (Fig.3b).
At the same time, the amount of delocalization measured by the
diffusion constant also increases.
For all the numerical results presented in this paper,
we use the Gaussian wave packet as the
initial condition.
However, we also checked that the qualitative
results given in this paper are insensitive to the initial condition.
This may be viewed as another characteristic of the chaotic system.
For QKR,
as the nonlinear parameter $K$ decreases, the results become more
sensitive to the initial condition \cite{Shio}.
In nonchaotic systems, the sensitivity in this sense can be used
to choose a preferred initial state in accordance to
some specific criterion, such as least entropy production \cite{ZHP}, etc.
 (For chaotic
cases, see \cite{Caves} for a related argument but from a different
point of view).

This may be explained in the following way.
Because we use a coupling through the momentum,
 the time scale for the system to lose coherence is given by
$ t_{dec} = {1 \over \gamma}(\frac{\lambda_{\theta dB}}{\delta p})^2 $,
where $\lambda_{tdB} = h / \sqrt{2\pi m k T}$
is the thermal de Broglie wavelength, and
 $\delta p$ is the relevant momentum scale.
After this time, noise will destroy the quantum coherence
between these momentum separation.
In the kicked rotor case, localization will occur due to the
coherence around $\delta p \sim  \Delta$, where $ \Delta \sim l\hbar $
 is the localization length.
Since $l \sim K^2$, this gives $ t_{dec} \sim {1 \over K^4} $.
Therefore, nonlinearity increases the rate of decoherence.

The relation between the diffusion constant $D$ and the noise strength
is given in \cite{OAH,CohFis}.
For our case, $K/\hbar \gg 1$ and for weak noises,
we can consider the particle as undergoing a random walk with
hopping parameter $1/t_c$.
Then $D = \frac{\Delta^2}{t_{dec}}$,  where $\Delta$ gives the localization
length. From this, we get
\begin{equation}
D = \gamma (\frac{\Delta^4}{\lambda_{\theta dB}^2})
\end{equation}

\section{Decoherence and Irreversibility}

The Wigner function
representation is often used to examine the quantum to classical
 transition.
For a linear system the Wigner function is known to show a smooth convergence
 to the classical Liouville distribution.
But if the system Hamiltonian has a nonlinear term,
quantum corrections associated with the higher derivatives of the
potential pick up the rapid oscillations in the Wigner function and it
no longer has a smooth classical limit \cite{BerHel}.
However, upon interaction with an environment,
a coarse-grained Wigner function can have a smooth
classical limit \cite{TakHab} for nonlinear systems

The Wigner function at time $t = j$ is defined as
\begin{equation}
W_{j}(X,p)
= \frac{1}{4\pi}\int\limits_{-2\pi}^{+2\pi} dy ~ e^{{i\over \hbar} py} ~
	\rho_{j}(X-{1\over 2}y, X+{1\over 2}y).
\end{equation}
where $X \equiv \frac{1}{2}(x+x'), y \equiv x-x'$.
 From (3.3), the unit time propagator for the Wigner function
 of QKR is,
\begin{equation}
W_{j+1}(X,p) =
 \exp[-\frac{K\sin X}{\hbar} \Delta_p] \exp[-\frac{p}{m} \partial_X]
			W_{j}(X,p)
\end{equation}
where $\Delta_p =  \exp[   \frac{\hbar}{2} \partial_p     ]
		 - \exp[  -\frac{\hbar}{2} \partial_p     ]
$ measures the effect of the kick.
We can see the effect of quantum corrections more clearly
if we expand $\Delta_p$ in orders
of $\hbar$:
\begin{equation}
\begin{array}{rl}
\Delta_p
 =  \exp[   \frac{\hbar}{2} \partial_p     ]
	- \exp[  -\frac{\hbar}{2} \partial_p     ]
 =  \hbar  \partial_p
	 +  \frac{\hbar^3}{24} \partial_ p^3
	 +  \frac{\hbar^5}{1920} \partial_ p^5
	 + ........
\end{array}
\end{equation}

With this, the first exponential in (4.2) contains the classical
propagator times quantum corrections of even orders of $\hbar$.
\begin{equation}
\begin{array}{l}
\exp [-\frac{K\sin X}{\hbar} \Delta_p]
		 =  \exp[    -K\sin X  \partial_p   ]
		\exp[  -\frac{\hbar^2}{24} K\sin X \partial_p^3  ]
		\exp[  -\frac{\hbar^4}{1920} K\sin X \partial_p^5]
		........
\end{array}
\end{equation}

If the initial system wavefunction is
discribed by a Gaussian wave packet
 with width $\delta p (>> \hbar$),
, we would expect to see a
classical-like evolution of the packet at short times.
When the width of the contracting wave packet gets small,
and becomes comparable to $\hbar$,
the effect of quantum corrections appears, namely, the corrections from
 the exponent which is higher order in $\hbar$ in (4.4).
By comparing the classical and the quantum terms, we can easily evaluate
the length scale at which quantum corrections become important,i.e. when
 $\delta p(t) \sim \hbar$.
Here $\delta p(t) = \delta p(0) e^{-\lambda t}$, where
$\lambda$ is the Lyapunov exponent given by $\lambda \sim ln (K/2)$ .
As shown in (1.2), from this expression, we can define the time scale $t_E$
as $t_E \sim ln \frac{\delta p(0)}{\hbar} / \lambda$.
Because the Wigner function or the expectation value of
any observable follows classical trajectories when $t < t_E$,
this has been called the Ehrenfest time.
Note that in the continuum case, this definition gives us a
different time scale for each term in the expansion \cite{ZurPaz}.
Hereafter, we set $m=1$ for brevity.

If the interaction with the environment has a form in
(3.6), the major effect of the bath is the appearance of a
 diffusion term in (4.2), such that,
\begin{eqnarray}
W_{j+1}(X,p) & = &
 \exp[D_{X} \partial_X^2]
 \exp[-\frac{K\sin X}{\hbar} \Delta_p]
 \exp[-p \partial_X]
			W_{j}(X,p)      \nn          \\
			 &  \approx &
 \exp[  D_{X} \partial_X^2 ]
 \exp[ -\frac{\hbar^3}{24} K\sin X \partial_p^3  ]
				W_{j}(X - p + K \sin X,p - K \sin X)
\end{eqnarray}
where $D_{X} = 2 M \gamma k T \hbar$ is related to a constant
 of the noise kernel
$\nu(s)$ defined before (2.16).

Competition amongst the three terms with different physical
origins is apparent:
The first term in (4.5) is the quantum diffusion term,
%\footnote{In the past litarature, two different term "diffusion"
%seem to exist.
% One is the $p^2$ diffusion in momentum space, the other is diffusion
% term responsible for the decoherence. In this case, we used in
% the second meaning. In this paper we tried to use the word decoherence
% for the second case in most cases just to avoid confusion.}
the second is the quantum correction term, and the third is purely
classical evolution.
As discussed by Zurek and Paz \cite{ZurPaz}, if D is sufficiently large,
the effect of quantum corrections becomes inconspicuous.
In this case, the diffusion term traces out a small scale oscillating behavior
before quantum corrections have a chance to  change the classical evolution.
Then one may expect the time evolution of the Wigner
 function to be like that of classical evolution with noise.
In this case, we can ignore the quantum correction in (4.5)
and write the evolution equation as
\begin{eqnarray}
W_{j+1}(X,p) & = &
	\exp[  D_{X} \partial_X^2 ]
	W_{j}(X - p + K \sin X,p - K \sin X)
\end{eqnarray}
The role of quantum diffusion is to add some Gaussian averaging
so that the contracting direction in phase space will be suppressed while
it does not affect the stretching direction.
As long as the width of the wave packet is large such that
the first term is negligible, the evolution should be Liouvillian
(time reversible if we assume infinite measurement precision).
Furthermore, we expect that after
the width of the packet along the contracting direction becomes comparable
to the diffusion generated width (in the Gaussian wave packet),
the dynamics will start showing irreversible
behavior arising from coarse graining (as distict from irrreversibility from
instability).
Consequently, entropy should increase in this regime.
In Fig. 4, we plot the von Neumann entropy for the dynamics of (4.6).
We can see three qualitatively different regimes:
I. the Liouville regime: the entropy is constant and the dynamics is
		time reversible.
II. the decohering regime: the entropy keeps increasing due to
		coarse graining.
III. the finite size regime: due to the bounded nature of the phase space,
		the entropy shows saturation.
Our result from quantitative analysis seems to confirm the qualitative
description of Zurek and Paz \cite{ZurPaz}
who used the inverted harmonic oscillator
potential as a generic source of instability.
Since the phase space in their model is not bounded
they do  not see Regime III.
Similar features appear in the quantum cap map (Fig. 5).
In this case, the full quantum dynamics can be calculated in a simple way.
Resemblance with the result of a classical rotor with noise is obvious.
However, in this case, the stable entropy
is smaller than the maximum value which may be explained as
a finite (phase space) size effect.\\

Quantum diffusion defined by the spreading of the
wavefuction is known to be dynamically stable.
The authors of \cite{Chi} performed the time reversal at some time and
saw the diffusion constant and even the wave function itself
 coming back to the same state within the accuracy of
computation.
As we know, these time reversal behaviors
cannot be seen in the classical case due to the instability of the
trajectories.
This is also true in real physical systems for which one can access
 information with only finite precision \cite{Ford}.

In Fig.6, we perform the time reversal at t = 200.
In the quantum case without bath, the system completely returns
to the original state after exactly the same amount of time.
Thus, the system is highly stable in spite of its random appearance.
On the other hand, in the classical case, instability prevents the
reversibility even without interaction with the bath.
When the interaction is turned on,
we see that the reversibility in the quantum system is gradually lost,
and irreversibility appears
as the noise strength increases.

In a realistic physical system which has a finite precision
 due to numerical or
instrumental limitation, we expect this type of irreversibity is
inevitable for the chaotic system even without noise.
If the minimum precision in length is denoted as $\epsilon$, the time
scale up to which the deterministic picture is valid is determined as
$t_p \sim \frac{1}{\lambda}log \frac{2\pi}{\epsilon}$.
At $t > t_p$, the system starts losing information about the past.
This may be the source of irreversibility for the classical chaotic system.
For the system we are studying, $t_p >> t_E > 1$ holds.
 Then the quantum effect smears the contracting evolution of the region in the
phase space before information about the past is lost.
Therefore, the wave packet traces back the same trajectory as it evolved.
If $t_c < t_p$, we see the irreversibility from coarse graining before
the limitation of the measurement becomes evident.
This type of irreversibility is another
characteristic of classicality peculiar to chaotic systems.\\

\noindent {\bf Acknowledgement}
We thank Drs. Shmuel Fishman and Juan Pablo Paz for explaining their work
and Drs. Ed Ott and Richard Prange for general discussions.
Research is supported in part by the National Science Foundation
under grant PHY91-19726. BLH also gratefully acknowledges support from
the General Research Board of the Graduate School of the University of
Maryland and the Dyson Visiting Professor Fund at the  Institute
for Advanced Study, Princeton.
A summary of this work was reported at the International
Symposium on Quantum-Classical Correspondence held at Drexel University
in Sept. 1994.
This work was completed while both authors enjoyed the hospitality of
the physics department of HKUST.

\newpage

\noindent {\bf Figure 1}
The linearized entropy (with opposite sign) is plotted against time.
If there is no environment, the entropy is constant for both the hyperbolic and
elliptic cases, indicating the purity of the state (dotted line).
For the hyperbolic map,
even though classically this system is strongly chaotic,
the corresponding quantum system does not show chaotic behavior.
This situation changes drastically when the system interacts with
a thermal bath.
In this case, the entropy due to coarse graining
keeps increasing.
Note that in the  hyperbolic case (solid line),
the rate of entropy increase is larger than in the elliptic case
 (dashed line).
 {\cal N = 50 } is used here (also in Fig. 2).

\noindent {\bf Figure 2}
The mean phase space point displacement is shown.
When there is no environment (dotted line), the system shows recurrence
in both the hyperbolic (a) and elliptic (b) cases.
In the presence of an environment (solid line),
 the hyperbolic map loses the recurrence
behavior under the Gaussian noise with $\sigma = 0.08$
and maintains a near-constant value, indicating
 ergodicity of the classical hyperbolic map.
On the other hand, the elliptic map still shows recurrence with the same
amount of noise, corresponding to the classical periodicity.

\noindent {\bf Figure 3}
$Tr \rho_r^2$ (solid line, left scale) and
$<p^2/2>$ (dashed line, right scale)  are plotted in Fig. 3a
versus time for $K = 10$ and $\hbar = 1$.
Three noise strength values
 $\sigma = 0$, $\sigma = 1.0$, $\sigma = 2.0$ are plotted here,
corresponding to the family of lines from up to down for $Tr \rho_r^2$ and
down to up for $<p^2/2>$ (note that $Tr \rho_r^2 = 1$ for $\sigma = 0$).
As the noise strength increases,
the decoherence time shortens, and
$Tr \rho_r^2$ decays rapidly.
This accompanies the increase of
diffusive behavior in $<p^2/2>$.
In Fig. 3b,
the same observables are plotted but with different $K$-values.
The upper, middle, and lower solid lines
(lower, middle, and upper dashed lines)
correspond
 to $K = 0.5$, $K = 5$, $K = 10$, respectively.
 Here $\sigma = 1.0$, $\hbar = 1$ are fixed.
We see that increasing nonlinearity shortens the decoherence time.
Note that when $K=0.5$, the system merely diffuses.

\noindent {\bf Figure 4}
The von-Neumann entropy is plotted versus time
for the kicked rotor.
For the first 25 steps, the system does not produce any entropy. The
evolution is reversible. Transition sets in in the next few steps,
the dynamics changes its character from reversible to irreversible.
Because the nonlinear term is from the sinusoidal function, the
onset of this regime is different at every phase space point.
Consequently we can only see the averaged behavior through
the entropy function.
Around $t = 40$,
 saturation occurs due to the finiteness and the periodic nature
of the phase space ($\sigma = 0.1$ for this case).

\noindent {\bf Figure 5}
The von-Neumann entropy is plotted versus time for the quantum cat map.
Due to the simplicity of the system, we see the same qualitative
features as in Fig. 4.
The entropy starts increasing around $t = 55$ and maintains a
near-constant rate of production while showing large oscillations.
After 200 steps, entropy production seems to saturate
and starts decreasing into some stable value.
 ${{\cal N} = 64 }$, $\sigma = 0.04$ are used here.

\noindent {\bf Figure 6}
The time reversal is performed at time = 200.
For QKR without a bath (lower curve), the system posseses time reversal
invariance \cite{Chi}.
 When the interaction with the environment increases,
 the system gradually regains
irreversibility as observed in classical systems.
Here the noise strength $\sigma = 0.5$ (middle curve)
 and $\sigma = 1.0$ (upper curve).
 Also $\hbar = 1$, $K = 10$.

\newpage

\end{document}